\documentclass[
aip,
reprint,%
amsmath,amssymb,
floatfix,
]{revtex4-1}

\usepackage{graphicx}
\usepackage{dcolumn}
\usepackage{bm}

\usepackage[utf8]{inputenc}
\usepackage[T1]{fontenc}
\usepackage{mathptmx}
\usepackage{etoolbox}
\usepackage{upgreek}
\usepackage{braket}
\usepackage[colorlinks=true,linkcolor=blue,anchorcolor=blue,citecolor=blue,urlcolor=blue]{hyperref}
\usepackage[final]{changes}

\makeatletter
\def\@email#1#2{%
 \endgroup
 \patchcmd{\titleblock@produce}
  {\frontmatter@RRAPformat}
  {\frontmatter@RRAPformat{\produce@RRAP{*#1\href{mailto:#2}{#2}}}\frontmatter@RRAPformat}
  {}{}
}%
\makeatother

\begin{document}


\title{Flopping-mode spin qubit in a Si-MOS quantum dot}

\author{Rui-Zi Hu}
\thanks{These authors contributed equally to this work.}
\author{Rong-Long Ma}
\thanks{These authors contributed equally to this work.}
\author{Ming Ni}
\author{Yuan Zhou}
\author{Ning Chu}
\author{Wei-Zhu Liao}
\affiliation{
  CAS Key Laboratory of Quantum Information, University of Science and Technology of China, Hefei, Anhui 230026, China
}
\affiliation{
  CAS Center for Excellence and Synergetic Innovation Center in Quantum Information and Quantum Physics, University of Science and Technology of China, Hefei, Anhui 230026, China
}
\author{Zhen-Zhen Kong}
\affiliation{
  Key Laboratory of Microelectronics Devices \& Integrated Technology, Institute of Microelectronics,\\Chinese Academy of Sciences, Beijing 100029, China
}
\author{Gang Cao}
\affiliation{
  CAS Key Laboratory of Quantum Information, University of Science and Technology of China, Hefei, Anhui 230026, China
}
\affiliation{
  CAS Center for Excellence and Synergetic Innovation Center in Quantum Information and Quantum Physics, University of Science and Technology of China, Hefei, Anhui 230026, China
}
\affiliation{
 Hefei National Laboratory, University of Science and Technology of China, Hefei 230088, China
}
\author{Gui-Lei Wang}
\affiliation{
  Key Laboratory of Microelectronics Devices \& Integrated Technology, Institute of Microelectronics,\\Chinese Academy of Sciences, Beijing 100029, China
}
\affiliation{
 Hefei National Laboratory, University of Science and Technology of China, Hefei 230088, China
}
\affiliation{
  Beijing Superstring Academy of Memory Technology, Beijing 100176, China
}
\author{Hai-Ou Li}
\thanks{Corresponding author: haiouli@ustc.edu.cn}
\affiliation{
  CAS Key Laboratory of Quantum Information, University of Science and Technology of China, Hefei, Anhui 230026, China
}
\affiliation{
  CAS Center for Excellence and Synergetic Innovation Center in Quantum Information and Quantum Physics, University of Science and Technology of China, Hefei, Anhui 230026, China
}
\affiliation{
 Hefei National Laboratory, University of Science and Technology of China, Hefei 230088, China
}
\author{Guo-Ping Guo}
\affiliation{
  CAS Key Laboratory of Quantum Information, University of Science and Technology of China, Hefei, Anhui 230026, China
}
\affiliation{
  CAS Center for Excellence and Synergetic Innovation Center in Quantum Information and Quantum Physics, University of Science and Technology of China, Hefei, Anhui 230026, China
}
\affiliation{
 Hefei National Laboratory, University of Science and Technology of China, Hefei 230088, China
}
\affiliation{
  Origin Quantum Computing Company Limited, Hefei, Anhui 230026, China
}





\date{\today}

\begin{abstract}
  Spin qubits based on silicon-metal-oxide-semiconductor (Si-MOS) quantum dots (QDs) are promising platforms for \deleted{a }large-scale quantum \replaced{computers}{computer}. 
 To control spin qubits in QDs, \deleted{the }electric dipole spin resonance (EDSR) \replaced{has been}{is} most commonly used in recent years. By delocalizing an electron across a double quantum dots (DQD) charge state, "flopping-mode"  EDSR has been realized in Si/SiGe QDs. Here, we demonstrate a flopping-mode spin qubit in a Si-MOS QD via Elzerman single-shot readout. When changing \added{the} detuning with a fixed drive power, we achieve s-shape spin resonance frequencies, an order of magnitude improvement in the spin Rabi frequencies, and virtually constant spin dephasing times. 
  Our results offer a route to large-scale spin qubit systems with higher control fidelity in Si-MOS QDs.
\end{abstract}


\maketitle 

Spin qubits in silicon QDs are a leading candidate for building a quantum processor due to their long coherence time~\cite{Dzurak2014AddressableQuantumDot, Dzurak2015TwoLogicGate}, potential scalability~\cite{Veldhorst2017InterfacingSpinQubits, Veldhorst2018CrossbarNetwork}, and compatibility with advanced semiconductor manufacturing technology~\cite{Camenzind2021FinFET, Vandersypen2022AdvancedSemiconductor, Zhang2018Semiconductor}. \replaced{Currently}{Nowadays}, as an alternative to implementing electron spin resonance (ESR)~\cite{Koppens2006ESR, Huang2019FidelitySilicon, Veldhorst2020HotSilicon, Chan2021ExchangeSilicon, Morello2022PrecisionTomography, Gilbert2022OnDemandControl}, EDSR allows the single-qubit and two-qubit operation fidelities to achieve 99.9\%~\cite{Dzurak2018SiliconSpinQubit, Tarucha2018CoherenceLimit} and 99\%~\cite{Vandersypen2022QuantumLogic, Tarucha2022FastUniversal, Mills2022TwoQubitPetta}, respectively, and allows for qubit operation at higher temperatures~\cite{Dzurak2020AboveOneKelvin}.


To implement EDSR in Si-MOS QDs, a rectangular micromagnet is deployed to generate an inhomogeneous magnetic field and an oscillating electric field resonant with the Larmor frequency is coupled to drive the spin states~\cite{Rashba2003OrbitalMechnisms, Leon2020MultiElectronSiMOS, Zhang2021Synthetic, Hu2021OperationGuide}. During the conventional EDSR measurement, electrons in Si-MOS QD are confined in the quantum well, leading to a relatively small electric dipole~\cite{Pioro2008EDSR, Kawakami2014SiGeEDSR}. Driving single spin rotations in a DQD close to zero detuning where electron shuttles between two QDs, the "flopping-mode" EDSR increases the electric dipole in QDs~\cite{Benito2019FloppingTheory}. A longer spin coherence time ($T_2^\text{Rabi}$) with the same Rabi oscillation frequency ($f_\text{Rabi}$) has been achieved in Si/SiGe spin qubits by applying flopping-mode EDSR via dispersive readout~\cite{Croot2020FloppingSiGe}. However,  the small size and complicated distribution of Si-MOS QDs make cavity readout of a flopping-mode spin qubit in Si-MOS QDs difficult~\cite{Dzurak2014AddressableQuantumDot, Huang2019FidelitySilicon, Veldhorst2020HotSilicon, Dzurak2020AboveOneKelvin, Leon2020MultiElectronSiMOS, Chan2021ExchangeSilicon, Gilbert2022OnDemandControl}.


Here, we demonstrate a flopping-mode single spin qubit in a Si-MOS QD via the Elzerman single-shot readout~\cite{Elzerman2004Single-shot}. By setting gate voltages carefully, a DQD with appropriate tunneling rates of an electron from \added{the }QD to \added{the }reservoir is formed underneath adjacent electrodes. Then, we measure the EDSR spectra, Rabi oscillation, and Ramsey fringes. Due to the large \added{interdot tunnel coupling }$2t_\text{c}$, an s-shape spin resonance frequency ($f_\text{q}$) as a function of the energy detuning ($\varepsilon$) is formed. We achieve an order of magnitude improvement in $f_\text{Rabi}$ around $\varepsilon=0$ with the spin dephasing times ($T_2^*$) and spin coherence time of Rabi oscillation ($T_2^\text{Rabi}$) virtually constant.


\begin{figure*}
  \includegraphics{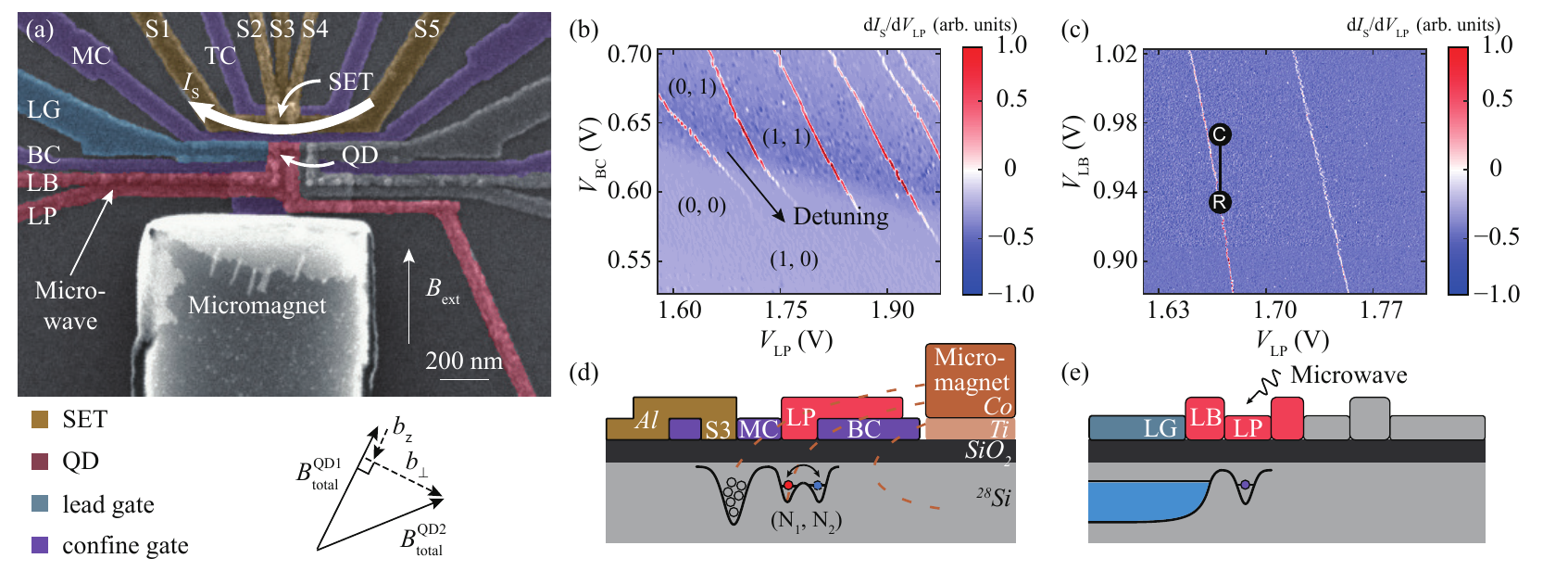}
  \caption{\label{fig:1} Flopping-mode spin qubit device layout and operation. (a) False-color SEM image of the device. The DQD is formed underneath \replaced{gates}{gate} LP (red) and BC (blue). The confine gates \added{TC, MC and BC }(purple) laterally confine the QDs and the single-electron transistor (SET). \added{The positions of QDs and SET are indicated by two white thin arrows. }The lead gate \added{LG }(blue) is the reservoir gate supplying electrons for the QDs. The SET is tuned by setting the voltages of the yellow gates \added{S1 to S5 }as a charge sensor of QDs. The white \added{bold }arrow above the SET indicates the SET current ($I_\text{S}$) direction. Gate LP (red) is connected to impedance-matched high-frequency lines via cryogenic bias-tees. \replaced{The}{And the} left white arrow indicates that \replaced{microwaves are}{microwave is} applied to gate LP. The right white arrow indicates the direction of the external magnetic field ($B_\text{ext}$). The cobalt (Co) rectangular micromagnet at the bottom of the image is used to generate an inhomogeneous magnetic field. The total magnetic field ($B_\text{total}$) contains $B_\text{ext}$ and the magnetization field of \added{the }micromagnet. The difference \replaced{in}{of} $B_\text{total}$ between two QDs can be divided into longitude ($b_\text{z}$) and transverse ($b_\bot$) components. (b) Charge stability diagram of the DQD \deleted{is }measured by differentiating $I_\text{S}$ as a function of gate voltages $V_\text{LP}$ and $V_\text{BC}$. The electron numbers in \added{the }QD underneath \replaced{gates}{gate} LP and BC are labeled $(N_1, N_2)$ on the diagram. The direction of the energy detuning ($\varepsilon$) in the DQD is indicated by the black dashed arrow. (c) Charge stability diagram of the DQD  as a function of gate voltages $V_\text{LP}$ and $V_\text{LB}$. We apply two-step pulse sequences to gate LB for the Elzerman single-shot readout. The relative voltage magnitude at each step of the pulse sequence for qubit manipulation is illustrated by the black line between two circles. During the measurement, we calibrate $V_\text{LB}$, $V_\text{LP}$ and $V_\text{BC}$ to maintain the tunneling rate of the electron from \added{the }DQD to the reservoir at the transition line. (d) Cross-sectional schematics of the device fabricated on a purified silicon-28 epi-layer. Electrons confined in the left quantum well underneath the SET gates \added{S3 }(yellow) are sensitive to charge movement in the QD region. The red ($N_1$) and blue ($N_2$) circles on the right side of the quantum wells represent the electrons in the flopping-mode regime. (e) Cross-sectional schematics of the device along the perpendicular direction. The electrons in the DQD tunnel across the barrier under gate LP to the reservoir under lead gate\added{ LG}. The microwave is applied to gate LP to rotate the electron spin, as mentioned in (a).}
\end{figure*}

Fig.~\ref{fig:1}(a) shows a scanning electron microscope (SEM) image of a typical Si-MOS DQD device, nominally identical to the one measured in Ref.~\cite{Hu2021OperationGuide}. The device was fabricated on a natural silicon substrate with a 70 nm thick isotopically enriched $^{28}$Si epi layer \replaced{that}{which} has a residual $^{29}$Si concentration of 60 ppm. \replaced{Overlapping}{The overlapping} aluminum gate electrodes were fabricated using multi-layer gate stack technology~\cite{Zhang2020GiantAnisotropy}.  The cobalt micromagnet integrated near QDs will be fully magnetized during the measurement, leading to a transverse magnetic field gradient for driving the spin qubits~\cite{Zhang2021Synthetic}. The total magnetic field at the QDs ($B_\text{total}$) is the sum of \added{the external magnetic field }$B_\text{ext}$ and the stray field from \added{the }micromagnet, as shown in the right bottom corner of Fig.~\ref{fig:1}(a). The device is in a dilution refrigerator at an electron temperature of $T_\text{e}=182.7\pm0.6$ mK (see Sec.~S2 in the Supplementary Materials for details).

The electrons are confined in the potential wells under gates LP and BC and form the DQD by selectively tuning gates LP, LB, and BC, as shown in Fig.~\ref{fig:1}(d). The corresponding charge stability diagram is shown in Fig.~\ref{fig:1}(b). $(N_1, N_2)$ on the diagram labels the corresponding number of electrons. \replaced{The}{And the} black arrow illustrates the direction of \replaced{$\varepsilon$}{energy detuning ($\varepsilon$)} between the DQD. \replaced{Gates}{Gate} MC and BC are designed to create a channel under the lead gate \added{LG }for electrons to tunnel between the electron reservoir and the DQD. Due to the small \replaced{electrode}{electrodes'} size ($\sim30$ nm) and the difference in \added{the }thermal expansion coefficient between the \replaced{aluminum}{Aluminum} electrodes and $\text{SiO}_2$ substrate surface, there is usually one quantum-well formed under each electrode gate in Si-MOS QDs, possibly forming complicated quadruple or more quantum dots in the device designed for the DQD.




Then, we apply two-step pulse sequences to gate LB for the Elzerman single-shot readout, as shown by points R (Read) and C (Control) in Fig.~\ref{fig:1}(c). The gate LB \replaced{is}{are} designed to modify the tunneling rate of electrons from \added{the }DQD to \added{the }electron reservoir, as shown in Fig.~\ref{fig:1}(e). Due to the capacitive coupling between gate LB, LP and BC, we need to calibrate the gate voltages $V_\text{LB}$, $V_\text{LP}$ and $V_\text{BC}$ to maintain the tunneling rate for different $\varepsilon$ during the measurements. \replaced{We confirm that}{And we confirm} the transitions between points R and C are adiabatic, as discussed in Sec.~S3 in the Supplementary Materials.

An external magnetic field is applied to the device for Elzerman single-shot readout. $B_\text{ext}$ is set to $605\ \text{mT}$ to induce \deleted{the }Zeeman splitting between spin states and fully magnetize the micromagnet magnetic\added{ field}. As a result, $\sim$20 GHz microwave pulses are applied to the LP gate via a cryogenic bias-tee to manipulate the qubit. By using sequences of selective EDSR pulses with microwave burst of frequency ($f_\text{s}$) at point C, we can perform single-qubit operations on the electron. The spin state is read out via state-to-charge conversion at point R, and a $\ket{\downarrow}$ electron is selectively loaded for initialization in the next pulse sequence~\cite{Hu2021OperationGuide}. The details of the measurement circuits are discussed in Sec.~S1 in the Supplementary Materials.

\begin{figure}[t]
  \includegraphics{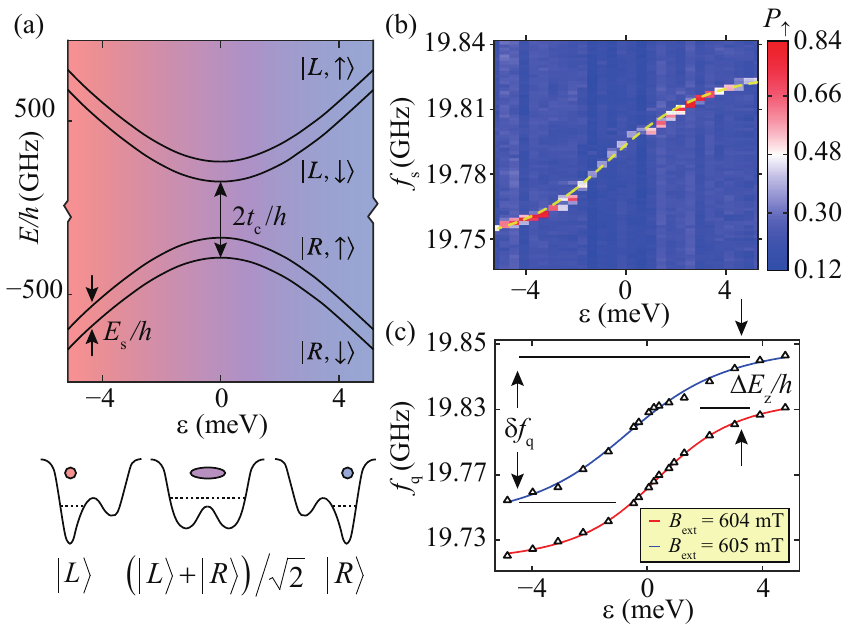}
  \caption{\label{fig:2} (a) Top: The eigenenergies calculated by diagonalizing the Hamiltonian in Eq.~\ref{eq:1}. Bottom: Schematics of \added{the }potential well for far detuned (left and right) and flopping-mode (middle) regimes. (b) EDSR spectra for the probability of $\ket{\uparrow}$ electrons ($P_\uparrow$) as a function of both $\varepsilon$ and microwave frequency ($f_\text{s}$). The dashed curve marks the position of increased $P_\uparrow$, which represents coarse ranges of spin resonance frequency ($f_\text{q}$). (c) $f_\text{q}$ as a function of $\varepsilon$ for $B_\text{ext}=604\ \text{mT}$ (red) and $B_\text{ext}=605\ \text{mT}$ (blue). The left arrows indicate the Zeeman energy difference ($\delta f_\text{q}$) generated by \added{the }longitudinal magnetic field difference ($b_\text{z}$) of \added{the }micromagnet. The right arrows illustrate the splitting energy difference ($\Delta E_\text{z}$) for different $B_\text{ext}$.}
\end{figure}

To detect the EDSR spectra rapidly, we apply frequency-chirped microwave pulses ($\pm$2 MHz around the center frequencies ($f_\text{s}$), 100 $\upmu$s duration times) to gate LP before the end of \replaced{the control}{Control} phase~\cite{Hu2021OperationGuide, Shafiei2013ChirpSiGe, Laucht2014ChirpPhosphorus, Sigillito2019ChirpSiGe}. If the frequency sweeps through \added{the }spin resonance frequencies $f_\text{q}$, the electron spin will end up in the excited state $\ket{\uparrow}$. By selectively setting $V_\text{LP}$, $V_\text{LB}$, and $V_\text{BC}$, we perform the Elzerman readout with a fixed tunneling rate of \replaced{approximately}{around} 150 Hz for the $\ket{\downarrow}$ electron to  ensure \deleted{the }consistency in the readout process while $\varepsilon$ increases from --4.5 to 4.5 meV along the exact transition line (0, 1) to (1, 0) (see Sec.~S2 in the Supplementary Materials for details). We measure the probability of \replaced{electrons}{electron} in the excited state ($P_\uparrow$) as a function of $f_\text{s}$ from 300 repeated single-shot readouts. For each $\varepsilon$, we repeat the measurement ten times, as mentioned in Ref.~\cite{Zhang2020GiantAnisotropy}. The EDSR spectra over $\varepsilon$ from --4.5 to 4.5 meV are shown in Fig.~\ref{fig:2}(b). There is an s-shape curve of increased $P_\uparrow$ with a \replaced{width}{wide} of 4 MHz of $f_\text{s}$, where $f_\text{q}$ is located. We calibrate the peak of $P_\uparrow$ and extract $f_\text{q}$ as a function of $\varepsilon$ in Fig.~\ref{fig:2}(c).

To explain this s-shape feature, we focus on the Hamiltonian $H$ of a single-electron occupied DQD system on the basis $(\ket{L\downarrow},\ \ket{L\uparrow},\ \ket{R\downarrow},\ \ket{R\uparrow})$~\cite{Benito2019FloppingTheory, Benito2017InputOutput}:
\begin{equation}
  H=\frac{1}{2}
  \left(
  \begin{matrix}
    -\varepsilon-E_\text{z1} & -2t_\text{SO} & 2t_\text{c} & 0\\
    -2t_\text{SO} & -\varepsilon+E_\text{z1} & 0 & 2t_\text{c}\\
    2t_\text{c} & 0 & \varepsilon-E_\text{z2} & 2t_\text{SO}\\
    0 & 2t_\text{c} & 2t_\text{SO} & \varepsilon+E_\text{z2}
  \end{matrix}
  \right).
  \label{eq:1}
\end{equation}
Here, $E_\text{z1}$ and $E_\text{z2}$ are Zeeman energies for the first and second QD, \added{respectively, }$2t_\text{c}$ is the interdot tunnel coupling, \added{and }$2t_\text{SO}=g\mu_\text{B}b_\bot$ is the synthetic spin-orbit coupling induced by \added{the }transverse magnetic field difference ($b_\bot$). 

The eigenenergies of this four-level system are shown in Fig.~\ref{fig:2}(a). The avoided crossings at $\varepsilon=0$ are generated by 2$t_\text{c}$. By diagonalizing the Hamiltonian in Eq.~(\ref{eq:1}), we calculate the energy splitting ($E_\text{s}$) between the lowest two energy levels as a function of $\varepsilon$. \added{Therefore, the corresponding spin resonance frequency of the qubit is obtained through $f_q\equiv E_s/h$. }For the situation of a small inhomogeneous field, i.e., $b_{\bot,z}\ll |\Omega-E_\text{z}|/g\mu_\text{B}$, where $\Omega=\sqrt{\varepsilon^2+4t_\text{c}^2}$, $E_\text{s}$ is corrected by the transverse and longitude \replaced{gradients}{gradient} to second and first order, respectively~\cite{Benito2019FloppingTheory}:
\begin{equation}
  E_\text{s}\simeq E_\text{z}-\frac{E_\text{z}^2-\varepsilon^2}{2E_\text{z}(\Omega^2-E_\text{z}^2)}(g\mu_\text{B}b_\bot)^2-\frac{\varepsilon}{\Omega}g\mu_\text{B}b_\text{z}.
  \label{eq:energy splitting}
\end{equation}
Here, $E_\text{z}=(E_\text{z1}+E_\text{z2})/2$ is the averaged Zeeman energy, \added{and }$\delta E_\text{z}=(E_\text{z1}-E_\text{z2})/2=g\mu_\text{B} b_\text{z}$ is the Zeeman energy difference generated by \added{the }longitudinal magnetic field difference ($b_\text{z}$) of the micromagnet. 

\begin{figure}[b]
  \includegraphics{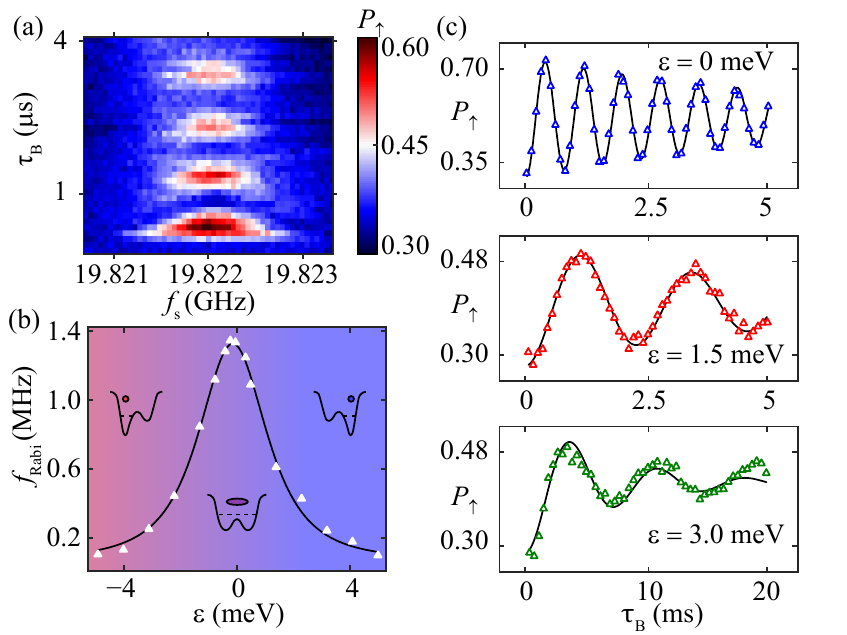}
  \caption{\label{fig:3} (a) The Rabi chevron is acquired at $\varepsilon = 0.5\ \text{meV}$. Every point of $P_\uparrow$ is obtained by repeating the pulse sequence 300 times with $\uptau_\text{B}$ fixed. We sweep the microwave frequencies, \deleted{then }repeat the measurement 10 times for every $\uptau_\text{B}$, and finally sum the results. (b) The Rabi frequency ($f_\text{Rabi}$) is plotted as a function of $\varepsilon$ with fixed microwave power. $f_\text{Rabi}$ is an order of magnitude larger at \replaced{approximately}{around} $\varepsilon=0$ than \added{at }the far detuned position. The solid curve is the fitting result of Eq.~\ref{eq:f_rabi}. (c) Rabi oscillations obtained at different $\varepsilon$. Top panel: $\varepsilon=0$ meV, $f_\text{Rabi}=1.262\pm0.002$ MHz and $T_2^\text{Rabi}=6.46\pm0.39$ $\upmu$s; Middle panel: $\varepsilon=1.5$ meV, $f_\text{Rabi}=0.429\pm0.003$ MHz and $T_2^\text{Rabi}=5.53\pm0.57$ $\upmu$s; Bottom panel: $\varepsilon=3.0$ meV, $f_\text{Rabi}=0.135\pm0.003$ MHz and $T_2^\text{Rabi}=7.01\pm0.82$ $\upmu$s.}
\end{figure}

We plot $f_\text{q}$ as a function of $\varepsilon$ for $B_\text{ext}=$ 605 and 604 mT in Fig.~\ref{fig:2}(c). By fitting $f_\text{q}$ with Eq~\ref{eq:energy splitting}, we obtain \replaced{$2t_\text{c}/h=914\pm167$ and $705\pm40$ GHz}{$2t_\text{c}=914\pm167$ and $705\pm40$ GHz} for $B_\text{ext}=$ 605 and 604 mT, respectively. The difference between the fitted splitting energy ($\Delta E_\text{z}=19.790\pm0.002-19.760\pm0.001$ GHz) equals the difference between the external magnetic fields ($g\mu_\text{B}\Delta B_\text{ext}/h=28$ MHz). $\delta f_\text{q}\equiv 2\delta E_\text{z}/h=56.8\pm4.5$ MHz for far detuned limits is shown in Fig.~\ref{fig:2}(c). 
In Ref.~\cite{Croot2020FloppingSiGe}, the lowest $E_\text{s}$ occurs near $\varepsilon=0$, leading to a sweet spot for spin dephasing. However, in our device $2t_\text{c}\gg E_\text{z}$, the second-order item $\frac{E_\text{z}^2-\varepsilon^2}{2E_\text{z}(\Omega^2-E_\text{z}^2)}(g\mu_\text{B}b_\bot)^2$ in Eq~\ref{eq:energy splitting} is suppressed, and there is no sweet spot \replaced{approximately}{around} $\varepsilon=0$.


After calibrating $f_\text{q}$, we now use a microwave burst with a specific burst time ($\uptau_\text{B}$) to manipulate the spin qubit. First, we measure $P_\uparrow$ as \replaced{a}{an} $f_\text{s}$ function with a fixed $\uptau_\text{B}$. Each point of $P_\uparrow$ in the curve is averaged from 300 repeated single-shot readouts. Then, we repeat the measurement ten times and sum $P_\uparrow$ with $\uptau_\text{B}$ changing from 0 to 4 $\upmu$s. The Rabi chevron is plotted in Fig.~\ref{fig:3}(a). 

Fig.~\ref{fig:3}(b) illustrates $f_\text{Rabi}$ as a function of $\varepsilon$ with fixed microwave power $P_\text{MW}=0$ dBm at the source. $f_\text{Rabi}$ is symmetric about $\varepsilon=0$ and is an order of magnitude larger at $\varepsilon=0$ than the far detuned position. For every $\varepsilon$, the corresponding $f_\text{Rabi}$ \replaced{is}{are} extracted by fitting \added{the }Rabi oscillation with the function $P_\uparrow(\uptau_\text{B})=A\cdot\text{exp}(-\uptau_\text{B}/T_2^\text{Rabi})\cdot\text{sin}(f_\text{Rabi}\uptau_\text{B})$, as shown in Fig.~\ref{fig:3}(c).

\begin{figure}[t]
  \includegraphics{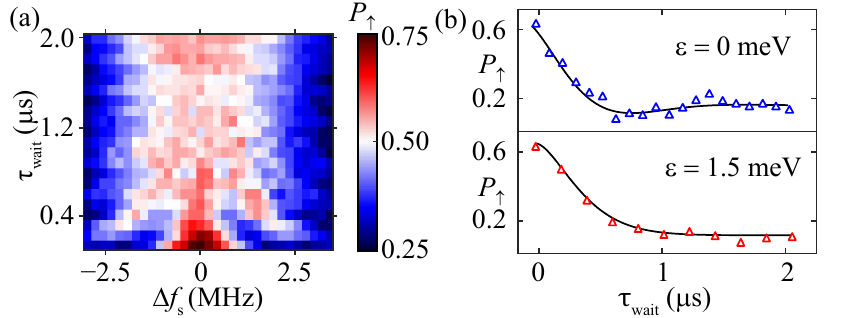}
  \caption{\label{fig:4} (a) Ramsey fringes as a function of frequency detuning ($\Delta f_\text{s}$) and waiting time ($\uptau_\text{wait}$) \deleted{is} measured at $\varepsilon = 0.5\ \text{meV}$ by applying a standard Ramsey fringe sequence with two $\pi/2$ pulses. Every point of $P_\uparrow$ is obtained by repeating the pulse sequence 300 times. We repeat the measurement ten times for every $\uptau_\text{wait}$ and sum the results. (b) The spin dephasing times ($T_2^*$) were measured through Ramsey fringes for different $\varepsilon$. Top panel: $\varepsilon=0$ meV, $T_2^*=0.42\pm0.31$ $\upmu$s; Bottom panel: $\varepsilon=1.5$ meV, $T_2^*=0.43\pm0.02$ $\upmu$s.}
\end{figure}

For a typical flopping-mode EDSR process, Ref~\cite{Benito2019FloppingTheory} gives $f_\text{Rabi}$ as a function of $\varepsilon$ for small $b_\bot$:
\begin{equation}
  f_\text{Rabi} = 4t_\text{c}^2g\mu_\text{B}b_\bot\Omega_\text{c}/\Omega|\Omega^2-E_\text{z}^2|.
  \label{eq:f_rabi}
\end{equation}
Here, $\Omega_\text{c}=edE_\text{ac}/\hbar$ is the Rabi frequency of charge qubits, proportional to the distance between the two QDs $d$, and the electric field with amplitude $E_{ac}$. $\Omega_\text{c}=15.8\pm0.8$ GHz can be obtained from the relevant result of $f_\text{Rabi}$ with a $g\mu_\text{B}b_\bot$ estimated as 0.232 $\upmu$eV~\cite{Zhang2021Synthetic}. We estimate $d\sim$ 0.02 $\upmu$m, thus $b_\bot\sim0.1$ T/$\upmu$m$\cdot 0.02$ $\upmu$m $=2$ mT.

Fig.~\ref{fig:3}(c) shows details of \added{the }Rabi oscillations for different $\varepsilon$. $f_\text{Rabi}=1.262\pm0.002$ MHz is achieved in the top panel. When $\varepsilon$ increases to 1.5 and 3 meV, the Rabi frequencies decrease to $f_\text{Rabi}=0.429\pm0.003$ and $f_\text{Rabi}=0.135\pm0.003$ MHz, respectively. By fitting \added{the }Rabi oscillation to an exponentially decaying sinusoid, $T_2^\text{Rabi}=6.46\pm0.39$ $\upmu$s at $\varepsilon=0$, $T_2^\text{Rabi}=5.53\pm0.57$ $\upmu$s at $\varepsilon=1.5\ \text{meV}$ and $T_2^\text{Rabi}=7.01\pm0.82$ $\upmu$s at $\varepsilon=3\ \text{meV}$ are obtained. $T_2^\text{Rabi}$ is stable when $\varepsilon$ increases. 

Furthermore, we measure $T_2^*$ for different $\varepsilon$ through Ramsey fringes. In Fig.~\ref{fig:4}(a), the Ramsey fringes are measured in the same way as the Rabi chevron. The averaged $T_2^*=0.42\ \upmu$s for $\varepsilon=$ 0 and 1.5 meV is acquired as shown in Fig.~\ref{fig:4}(b). As mentioned in Ref.~\cite{Zhang2021Synthetic}, \added{the }longitudinal magnetic field difference $b_\text{z}$ is one of the most relevant sources for dephasing in our device.
Since the sweet spot is absent and $b_\text{z}$ is constant during the measurement, we attribute the enhancement of the quality factor ($Q\equiv2T_2^\text{Rabi}f_\text{Rabi}$) to the improvement of the electric dipole.

In summary, we demonstrate the flopping-mode EDSR in a Si-MOS quantum dot through the Elzerman single-shot readout. We construct a DQD with \replaced{$2t_\text{c}/h\sim800$ GHz}{$2t_\text{c}\sim800$ GHz} under adjacent electrodes by selectively setting gate voltages. We extract an s-shape $f_\text{q}$ as a function of $\varepsilon$ from the EDSR spectra. Then, we improve $f_\text{Rabi}$ \added{by }an order of magnitude from $0.135\pm0.003$ to $1.262\pm0.002$ MHz by increasing the electric dipole. \deleted{And }$T_2^*$ and $T_2^\text{Rabi}$ remains \replaced{at approximately}{around} $0.42\pm0.03$ $\upmu$s and $6.46\pm0.39$ $\upmu$s, respectively. 
We anticipate that flopping-mode EDSR will have better performance in the heavy hole regime~\cite{Mutter2021HeavyHoleFlopping} or phosphorus donor qubits~\cite{Krauth2022PhosphorusFlopping}, and will perform two-qubit operation~\cite{Cayao2020FloppingTwoQubits} in the future.

\section*{Supplementary materials}
See the supplementary material for the description on the device, measurement setup, background subtraction and simulation in more detail. The larger charge stability diagram and the method for extracting the coupling strength are also presented.

\begin{acknowledgments}
This work was supported by the National Natural Science Foundation of China (Grants No. 12074368, 92165207, 12034018 and 61922074), the Innovation Program for Quantum Science and Technology (Grant No. 2021ZD0302300), the Anhui Province Natural Science Foundation (Grants No. 2108085J03), and the USTC Tang Scholarship. This work was partially carried out at the USTC Center for Micro and Nanoscale Research and Fabrication.
\end{acknowledgments}


\section*{References}
\bibliography{Flopping_mode_manuscripts}

\end{document}